# Direct Imaging of Graphene Edges: Atomic Structure and Electronic Scattering

*Jifa Tian[1,2,\*], Helin Cao[1,2], Wei Wu[3], Qingkai Yu[3,#], Yong P. Chen[1,2,4,\*]*

1. Department of Physics, Purdue University, West Lafayette, Indiana 47907

2. Birck Nanotechnology Center, Purdue University, West Lafayette, Indiana 47907

3. Center for Advanced Materials and Department of Electrical and Computer Engineering, University of Houston, Houston, Texas 77204

4. School of Electrical and Computer Engineering, Purdue University, West Lafayette, Indiana 47907

[#]Present Address: Ingram School of Engineering and Materials Science, Engineering and Commercialization Program, Texas State University, San Marcos, Texas 78666

*To whom correspondence should be addressed: E-mail: tian5@purdue.edu; yongchen@purdue.edu

ABSTRACT We report an atomically-resolved scanning tunneling microscopy (STM) investigation of the edges of graphene grains synthesized on Cu foils by chemical vapor deposition (CVD). Most of the edges are macroscopically parallel to the zigzag directions of graphene lattice. These edges have microscopic roughness that is found to also follow zigzag directions at atomic scale, displaying many ~120° turns. A prominent standing wave pattern with periodicity ~$3a/4$ ($a$ being the graphene lattice constant) is observed near a rare-occurring armchair-oriented edge. Observed features of this wave pattern are consistent with the electronic intervalley backscattering predicted to occur at armchair edges but not at zigzag edges.

KEYWORDS  Graphene, STM, edge, zigzag, armchair, electronic scattering

MANUSCRIPT TEXT. Graphene edges are fundamentally interesting[1,2] and can display rich behavior dependent on their atomic arrangements[3-7]. The two most basic configurations of graphene edges, "zigzag" and "armchair", are defined in terms of their orientations along graphene's two major crystallographic directions. These two prototypes of graphene edges have been predicted to have very different electronic properties that could be important for nanoelectronic applications. For example, theories predict that zigzag edges exhibit a characteristic edge state (absent in armchair edges) that can give rise to novel electronic and magnetic properties of graphene nanoribbons (GNRs) with zigzag edges[3-6]. The electronic scattering at armchair and zigzag edges are also markedly different. It has been suggested that armchair edges give rise to intervalley backscattering, which is suppressed at zigzag edges[7]. While edges that display *macroscopic* orientations parallel to either zigzag or armchair crystallographic directions are commonly observable, for example, in exfoliated graphene sheets[8,9], such (macroscopically-oriented) zigzag or armchair edges are generally expected to have microscopic edge *roughness* (deviation from perfect zigzag or armchair edges), which can have important implications for experiments and applications involving graphene nanostructures[10].

Scanning tunneling microscopy (STM) is a powerful tool to characterize the local structural and electronic properties of graphene down to atomic scale. However, it has been challenging to perform atomically-resolved STM imaging of edges of a single layer graphene to clearly demonstrate their atomic structures and orientation dependent electronic properties[9]. In this report, we present atomically resolved STM images showing unprecedented details of the atomic structure and roughness of zigzag edges in chemical vapor deposited graphene single crystal grains grown on Cu foils. We also observe a striking standing wave pattern parallel to an armchair edge, with features consistent with the electronic intervalley backscattering expected to occur at armchair but not zigzag edges.

Our studies were performed on hexagonally-shaped graphene grains (with typical size of several microns or bigger) synthesized *ex-situ* by ambient chemical vapor deposition (CVD) on Cu foils[11] (see Supporting Information). Our previous work[11,12] has shown that such grains are monolayer single crystalline graphene, with edges predominantly exhibiting macroscopic orientations parallel to zigzag directions. In the current work, such graphene grains as-grown on Cu substrates are placed in an UHV (ultrahigh vacuum) STM system and annealed at ~400 $^{o}$C for 48 hours before STM is performed at room temperature (see Supporting Information).

Figure 1a shows a large-scan-area STM topographic image taken near two edges (highlighted by dashed lines, forming an angle of ~120°) of one such graphene grain. The atomically resolved image (inset) acquired from inside the grain (from the area indicated by dotted black box in Figure 1a) exhibits characteristic honeycomb lattice of single layer graphene. From such a lattice image we infer the two edges in Figure 1a are (macroscopically) oriented along zigzag directions. Figure 1b is a zoomed-in image showing one of the edges (from the area indicated by the dotted red box in Figure 1a). This edge, with its macroscopic orientation parallel to one ($Z_1$) of the 3 zigzag directions (labeled in the image) of graphene, clearly displays microscopic roughness. Figures 1c and 1d are higher resolution STM images of two representative sections of this edge, showing striking details of its atomic arrangement (see also another example in Figure S1a). The edge is observed to consist mostly of shorter segments (with observed lengths ranging from a few to tens of lattice spacing) of atomically smooth zigzag configuration. The orientations of these locally-perfect zigzag segments can be parallel or 120° from the "global" orientation ($Z_l$) of the edge. Therefore, the edge takes a distinctive "zigzag-roughened" shape and displays many ~120° turns, while maintaining both the local and global zigzag orientations. Such "zigzag-roughened" edge structure is found in all the zigzag edges (the predominant edge types in our samples) we have imaged.

We note that the relative stability of zigzag versus armchair edges in graphene has been theoretically debated[13-16], with different conclusions reached in different calculations. Experimentally, both armchair and zigzag oriented edges (which may still have microscopic roughness) have been found in graphene prepared by different methods. Exfoliated graphene flakes appear to show zigzag and armchair edges with comparable likelihood [8,9,17]. On the other hand, graphene grains grown on various metal surfaces[18-20] appear to prefer zigzag edges, while those grown on SiC (0001) appear to prefer armchair edges[21], suggesting that the stability of zigzag versus armchair edges may depend on various factors, such as graphene's local environment (particularly substrates). Both the global[11] and local orientations of the edges of our graphene grains are found to be predominantly parallel to zigzag directions, suggesting that zigzag edges are strongly preferred in this system. The ubiquitous occurrence of zigzag-oriented edges we observed, spanning a wide range of length scales (from microns to angstroms, Figures. 1 and S1), may shed new light on the structures of graphene edges, as well as provide insights on the growth mechanisms and kinematics of CVD graphene on Cu.

Although the dominant majority of the edges in our graphene grains are found to be zigzag-oriented[11], we did observe one rare-occurring armchair-oriented edge (labeled by the dotted line in Figure 2a, forming ~30° angle with two nearby zigzag-oriented edges labeled by dashed lines). This is so far the only armchair edge we found in our STM experiments, despite our efforts to search for more. Figure 2b shows a 3D rendition of a high resolution STM topographic image acquired from the area indicated by the dotted red box in Figure 2a. We observe a striking standing wave like pattern parallel to the armchair edge (dotted white line, where the wave pattern has made it difficult to image the armchair edge itself at atomic scale). In contrast, such a wave pattern is not observed (within our imaging resolution) near the zigzag edge in Figure 2b (dashed white line) or any other zigzag edges we have imaged (Figures. 1 and S1). Figure 2c shows a representative atomically resolved image taken near the armchair edge, showing the periodicity ($\lambda$) of the wave pattern (dashed black lines) is approximately *3a/4, a* being the graphene lattice constant (see Supporting Information). This periodicity corresponds to a wavevector of *2π/λ~8π/3a*. Figure 3a shows the Fourier transform (FT) of Figure 2c, where the FT of the wave pattern gives rise to the two spots labeled by the red circles. These two spots can be identified with the "*2K*" points in the momentum-space model of graphene[22] in Figure 3b (which also shows the Dirac "*K*"-points as labeled by dashed red circles and the reciprocal lattice (RL, FT of graphene lattice) points, as labeled by solid black circles and corresponding to the 6-fold-symmetric spots in Figure 3a. The extracted length ratio between the vectors "*RL*" and "*2K*" labeled in Figure 3a is ~*0.82*. The extracted ratio ($\lambda/a$) between the periodicity ($\lambda$) of the observed wave pattern and graphene lattice constant (*a*) is *λ/a~0.73* (by averaging over different areas of the real space STM images, Figure 2c and Figure S2a). Both these values are in excellent agreement with the theoretical value (21) of $RL/2K = \sqrt{3}/2$ (~0.866) and *(2π/2K)/a=3/4* (where *2π/2K* is the wavelength corresponding to a wavevector *2K*), within our experimental uncertainty (mainly due to tip scan hysteresis and sample/tip drift resulting in a slight distortions in our STM images).

It is well known that STM is sensitive to the local electronic density of states (DOS) and standing wave patterns resulted from electronic scattering are often revealed (even at room temperatures) in STM topographs[23-25]. While more work is needed to confirm whether the standing wave pattern observed at the sole armchair edge we found in our samples is a generic phenomenon for armchair edges in graphene, several features including the direction (normal to the armchair), periodicity (~*3a/4*) and wavevector (~*2K=8π/3a*) of the standing wave pattern shown in Figure 2 are all consistent with a predicted intervalley backscattering induced by the armchair edge[7]. Such an intervalley backscattering connects two

Dirac cones (Figure 3c, "*K*" and "-*K*", where the length of the "*K*" vector is the *ΓK* distance = *4π/3a*) [22] oppositely located from *Γ* point in the momentum space (Figure 3b), corresponding to a scattering wavevector of *2K*. Several Raman spectroscopy studies[7,9,26] of graphene edges have suggested[7] that an armchair edge (terminated by both "*A*" and "*B*" sub-lattice sites, Figure 3d) induces strong intervalley backscattering, as the backscattering direction (normal to the edge) is well aligned with an intervalley scattering wavevector (*2K*). In contrast, intervalley backscattering is suppressed at a zigzag edge (terminated by "*A*" sites only) as such an alignment does not occur (instead the backscattering direction makes a 30° angle with a *2K* vector). In our experiments, the standing wave pattern shown in Figure 2 was only observed at this armchair edge. It was not observed at any of the (many) zigzag edges we studied nor in any other regions inside graphene grains or on Cu surfaces (including at Cu surface steps), thus unlikely to be associated with some random topography features (unrelated to armchair edges) of graphene or Cu surfaces. It is also unlikely to be related to the "$2k_F$" Friedel oscillations of Cu surface states, observed previously by low temperature STM on single crystalline Cu, eg. at surface steps [24,25], with an oscillation periodicity (~1.5 nm) much larger than ours. Taken together, these considerations are highly suggestive that the observed standing wave pattern is related with the intervalley backscattering expected to occur at the armchair edge of graphene. STM imaging can thus provide a clear demonstration of the remarkable difference in the electronic scattering properties between armchair and zigzag edges previously suggested by Raman studies.

Recently, a scattering pattern with wavevector close to *K* (periodicity ~*2π/K=3a/2*) has been observed by STM near armchair-oriented edges of graphene grown on SiC (0001) [23]. The authors [23] also related this scattering pattern with intervalley backscattering, and attributed the apparent "doubling" of the observed periodicity (from the expected *2π/2K* for intervalley backscattering) as due to the localization of electronic DOS along the C-C bonds in graphene, resulting in the "missing" of every other oscillation. In comparison, our observed scattering pattern has wavevector ~*2K* (periodicity ~*2π/2K*), as expected for intervalley backscattering. We suggest that this difference may be related to the different substrates, which can have important influence on electronic properties of graphene. In our case, the metallic Cu substrate may have played some role to enable the full oscillations of intervalley backscattering in graphene (periodicity *2π/2K*, half that observed in Ref. 23 to be observed. More work is needed to understand this effect in detail. We note that no clear observation of intervalley backscattering patterns have been obtained in previous STM measurements on armchair edges of *nanometer-sized* graphene islands[17, 21], probably due to the obscuring effect of interfering multiple scatterings from edges of different directions in such small size systems. The short "propagation distance" (on the order of a few nm's, beyond which the oscillation is no longer discernible) of the wave pattern we observed (Figures. 2b, 2c) may be generally attributed to thermally induced dephasing of electrons in such room temperature measurements[23].

Chemical vapor deposition (CVD) of graphene on Cu foils has emerged as one of the most promising routes to scalable production of graphene for a wide range of applications[27,28]. STM studies can be highly valuable to understand the growth mechanism of such CVD graphene and its structural and electronic properties at atomic scale. Such a system is also particularly convenient to perform fundamental STM studies of graphene as single layer graphene can be easily synthesized on Cu substrates, which only weakly interacts with (and perturbs) graphene[11,29] while providing a highly conductive substrate facilitating STM of graphene and its edges.

Atomically-resolved images of graphene edges and edge roughness such as those shown in this work (Figures. 1 and S1) could provide a precise characterization of the disorder due to edge roughness that can strongly affect the behavior of graphene nanoelectronic devices[10]. It is interesting to note that, despite their microscopic roughness, the zigzag-oriented edges (whose global and local orientations both follow zigzag directions) of our graphene grains do not appear to scatter electrons strongly, enabling the atomic structure of such edges to be well resolved without being masked by scattering patterns (in contrast to the armchair edge in Figure. 2). Such "zigzag roughened" edges (largely free of electronic scattering) found in our CVD graphene on Cu have not been reported in previous STM studies of the edges in other forms of stand-alone graphene[9,17-21] or graphene-on-graphite[30-32] and may possess distinctive electronic properties and promise for device applications.

ACKNOWLEDGMENT This work was performed under the auspices of Argonne National Laboratory (ANL) Center for Nanoscale Materials (CNM) User Research Program (Proposal ID 998), and partially supported by NSF, DHS and NRI-MIND center. The user facilities at ANL's CNM are supported by the U. S. Department of Energy, Office of Science, Office of Basic Energy Sciences, under Contract No. DE-AC02-06CH11357. The authors thank N. P. Guisinger, L. Gao, J. Cho and J. R. Guest of CNM at ANL for experiment assistance and discussions.

FIGURE CAPTIONS

**Figure 1.** (a) STM topographic image of a graphene grain near 2 zigzag-oriented edges (dashed white lines). Measurement condition: tunneling current I=0.1 nA, sample-tip bias voltage V=-0.2 V. The lines texture (spacing ~ 4.5 nm) seen on graphene is attributed to features on Cu surface underneath graphene and not the focus of this paper. Inset (scale bar =2.4 Å) shows representative atomically resolved image of graphene lattice inside the grain taken from the area indicated by the black dotted box (I=20 nA, V=-0.2 V). (b) A magnified image, acquired from the red dotted boxed area in (a), showing the graphene edge with microscopic roughness (I=20 nA, V=-0.2 V). Arrows indicate the 3 zigzag directions in graphene lattice. (c and d) Representative atomically-resolved images of the graphene edge taken from the blue dotted boxed area in (a) and black dashed boxed area in (b), respectively, showing its roughness typically follows zigzag directions (I=20 nA, V=-0.2 V). A few model hexagons of graphene lattice are superimposed on the image. Dashed white lines mark the general contour of the edge. The Cu surface not covered by graphene is oxidized[33] and the topographic depression (dark area) just outside graphene (within ~1 nm from the graphene edge) is interpreted as due to less coverage of Cu oxide immediately near graphene.

**Figure 2.** (a) STM topographic image showing 2 zigzag-oriented edges and a rare-occurring armchair-oriented edge (I=0.1 nA, V=-0.2 V). Magnified images of the dotted red and green boxes are presented in (b) and Figure S2a, respectively. (b) A 3D rendition of atomic-resolved topographic image acquired from the red box (15.4 nm by 15.4 nm) in (a), showing a standing wave pattern near the armchair edge (I=10 nA, V=-0.2 V). Dotted and dashed lines mark the armchair ("A", whose precise location is obscured by the standing wave pattern) and zigzag ("Z") edges. (c) A zoomed-in image near the armchair edge (I=10 nA, V=0.2 V), with superimposed models for part of the graphene lattice (hexagons, lattice constant $a$) and the standing wave pattern (dashed lines, period $\lambda$).

**Figure 3.** (a) Fourier transform (FT) of image Figure 2c, showing reciprocal lattice (RL) points of graphene and two extra spots (red circles) associated with intervalley "$2K$" scattering. (b) Schematic of graphene reciprocal lattice (black circles), Dirac "$K$" points (dashed red circles), and two intervalley scattering ("$2K$") vectors (solid red circles). (c) Schematic of intervalley scattering process between two Dirac cones located at $K$ and $-K$. (d) Schematic of an armchair edge (terminated by both $A$ and $B$ sub-lattice sites) and a zigzag edge (terminated by $A$-sites only). The figure is oriented to correspond to the graphene $K$-space model shown in (b). Two dashed arrows indicate directions normal to the armchair and zigzag edges, respectively. The normal ($N_A$) to the armchair edge is oriented along the $2K$ vector shown in (b) and is also the propagation direction of the intervalley backscattering (modeled by the dashed lines) from this edge.

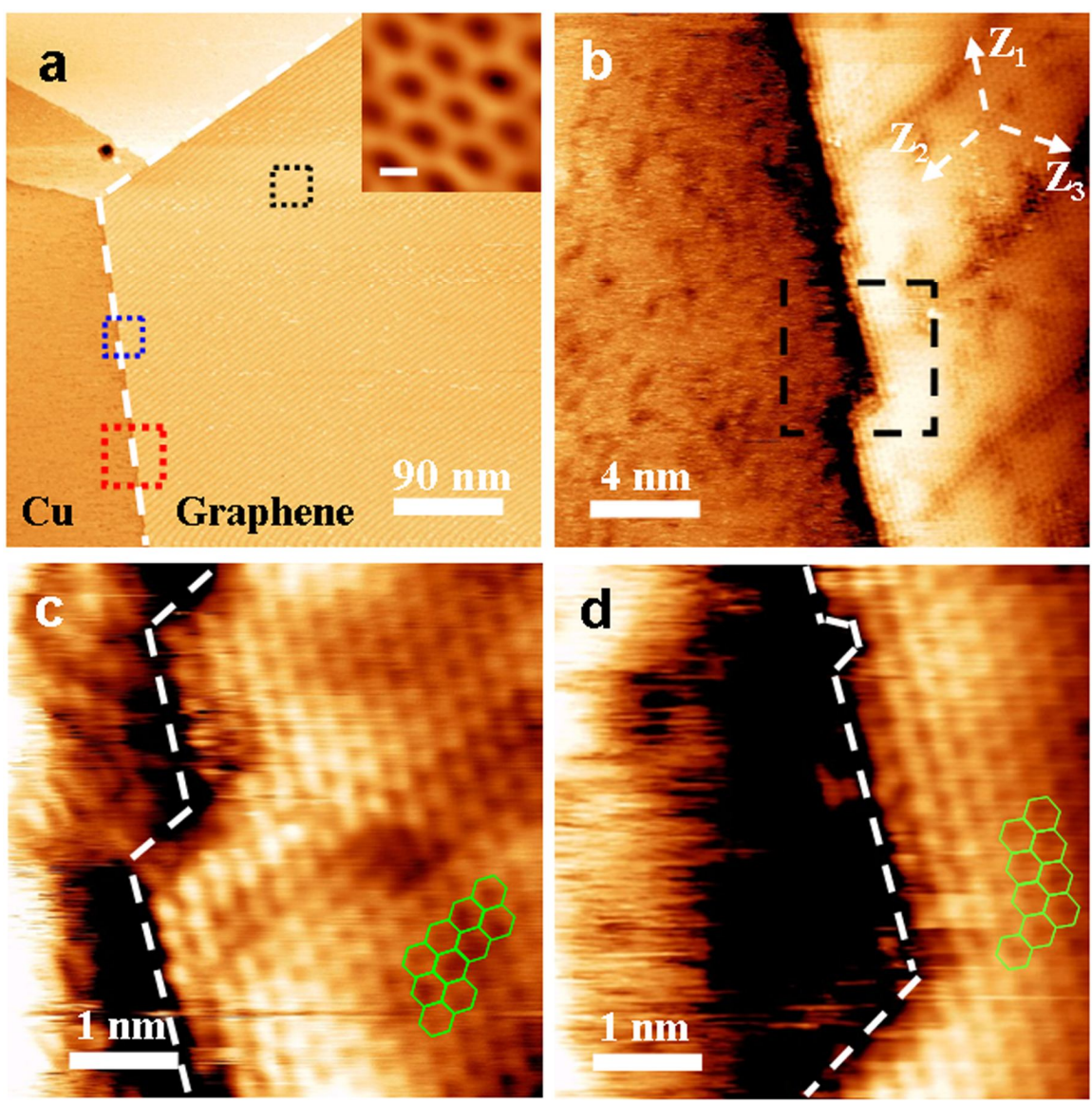

Figure 1 Jifa Tian *et al.*

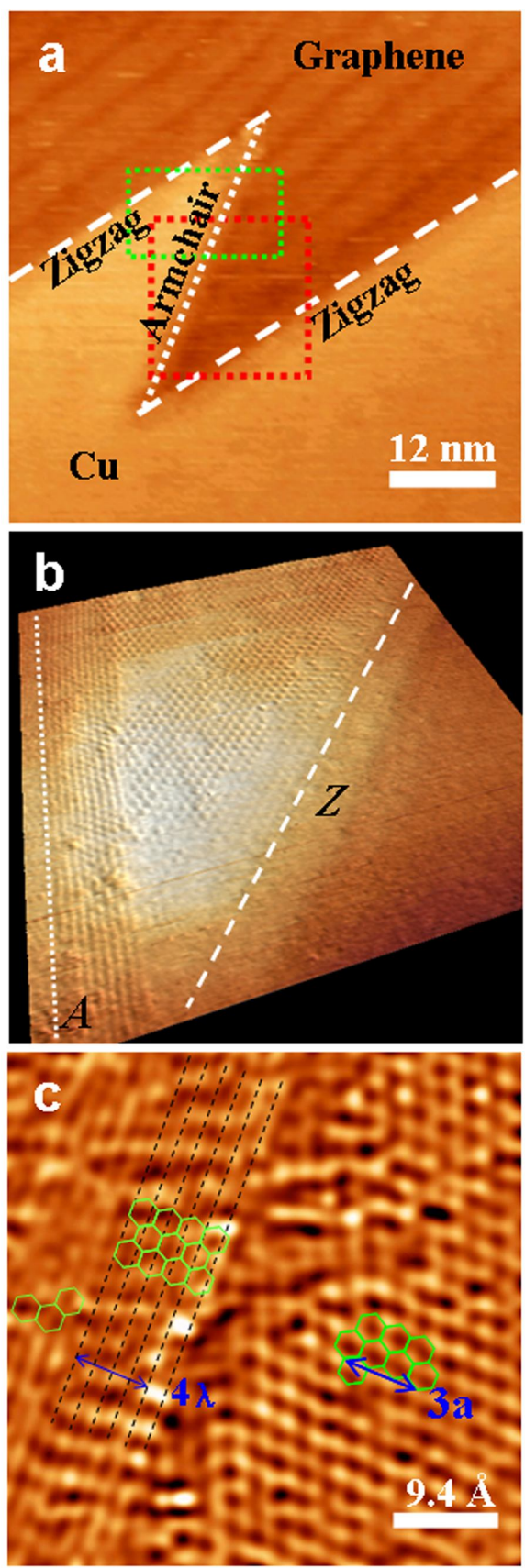

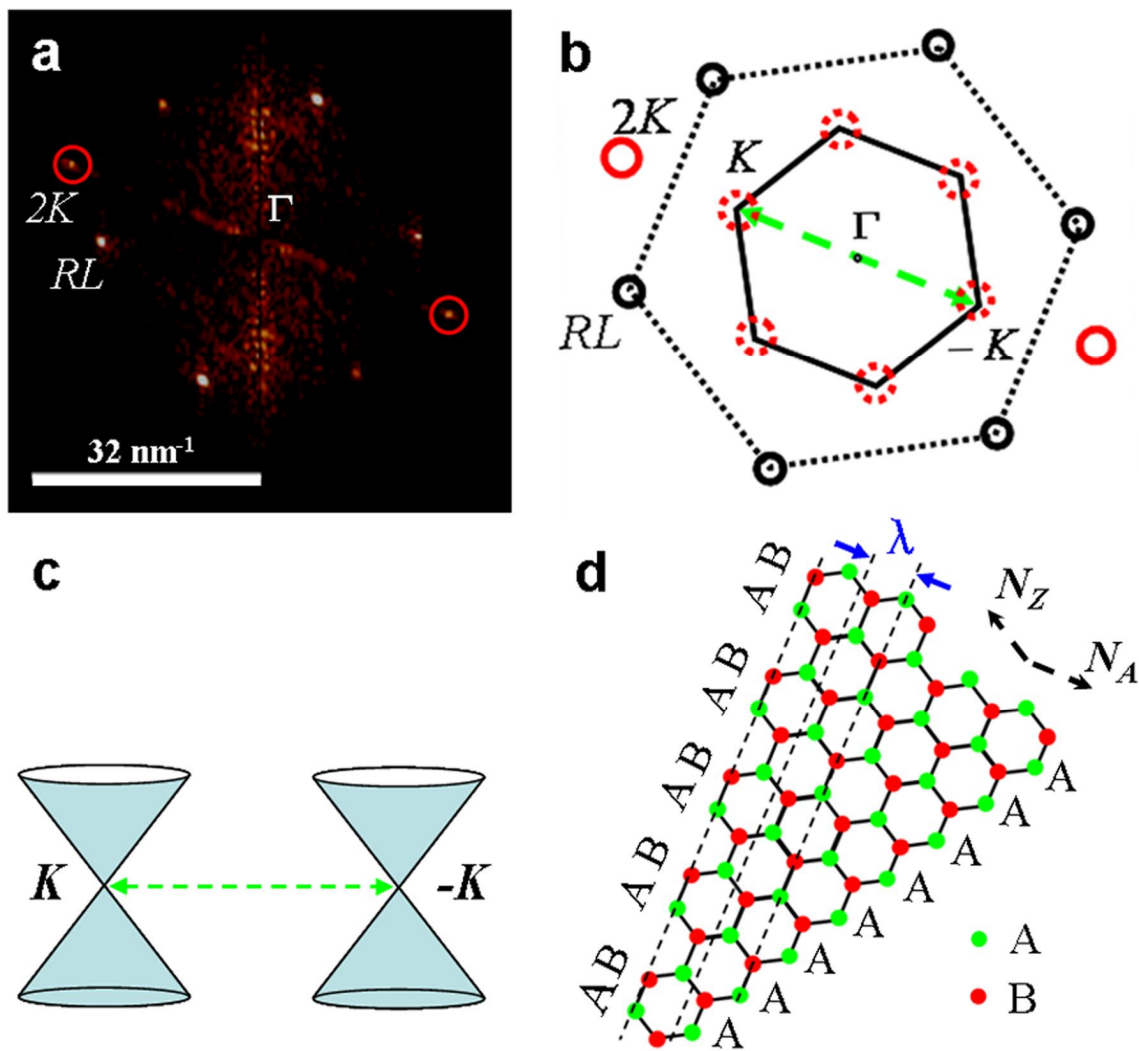

Figure 3 Jifa Tian *et al.*

# Supporting Information

**Materials and methods**

Graphene grains were grown by CVD ($CH_4$ as carbon stock) on Cu foils at ambient pressure[S1]. First, a Cu foil (25-μm-thick, 99.8%, Alfa Aesar) was loaded into a CVD tube furnace and heated up to 1050 ºC under 300 sccm Ar and 10 sccm $H_2$. After reaching 1050 ºC, the sample was annealed for 30 min without changing the gas flow rates. The growth was then carried out at 1050 ºC under a gas mixture of 300 sccm diluted (in Ar) $CH_4$ (concentration 8 ppm) and 10 sccm of $H_2$. The growth time was kept short (10-15 min) so that the graphene grains have not merged to form a continuous film[S1]. Finally, the sample was rapidly cooled to room temperature under the protection of Ar and $H_2$. After the growth, the sample (graphene grains on Cu) was moved into an Omicron ultra-high vacuum (UHV) STM system and annealed at 400 ºC for 48 hours. STM measurements were then carried out at room temperature in UHV (base pressure < $10^{-11}$ mbar) using electrochemically etched STM tips made of tungsten or platinum/iridium alloy. Differential conductance (dI/dV) maps were obtained by recording the scanning tunneling spectroscopy (STS) spectrum (using lock-in detection, with 30 mV rms amplitude modulation at ~ 10 kHz applied to the bias voltage) at each spatial pixel in the topography measurement.

**Supplementary Text**

Figure S1a shows another example of atomic resolved STM topographic image acquired from a zigzag-oriented graphene edge. One again observes that the atomic-scale roughness to always take zigzag directions. Fig. S1b shows a lower-resolution STM topographic image of another graphene edge (macroscopically zigzag oriented) highlighted by dotted white line, demonstrating larger-scale (~10nm) roughness (along zigzag directions, with many 120° turns).

Figure S2 shows another atomically-resolved STM topographic image (a) and corresponding dI/dV map (b) near the rare-occurring armchair-oriented edge studied in main Fig. 2. The images (acquired from the area corresponding to the dotted green box in Fig. 2a) also show one of the nearby zigzag-oriented edges. Near and parallel to the armchair edge, we again see the wave pattern (with periodicity ~3/4 of the graphene lattice constant) consistent with intervalley backscattering from the armchair edge. In contrast, no such wave pattern is observed near the zigzag edge.

References for Supporting Information:

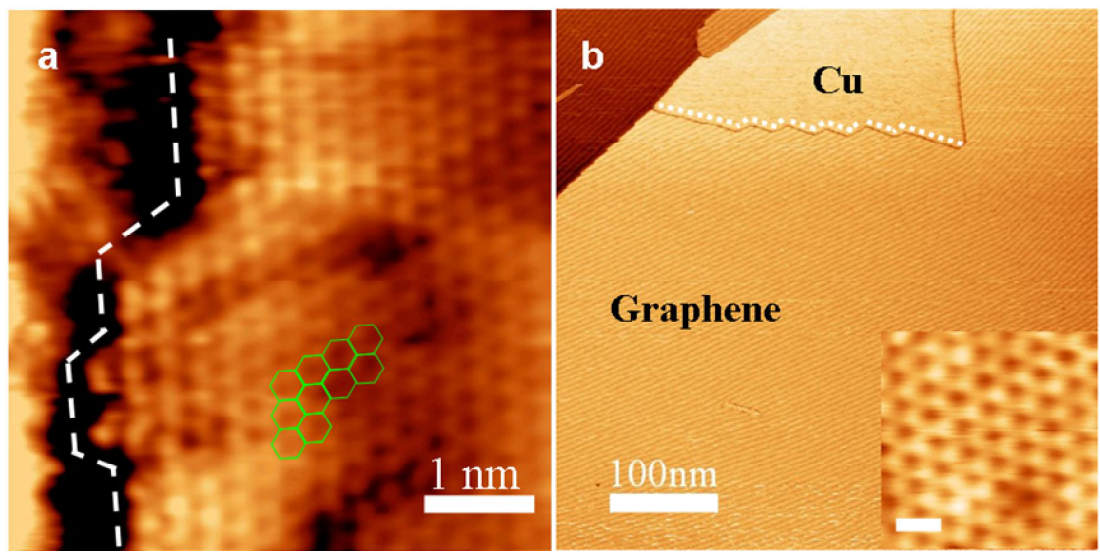

**Figure S1** (a) Atomically resolved image acquired from another zigzag oriented edge, demonstrating that the atomic-scale roughness largely follows zigzag directions. Data measured with tunneling current I=20 nA, sample-tip bias voltage V=-0.2 V. (b) Lower-resolution STM topographic image (I=0.1 nA, V=-0.2 V) of a graphene grain showing a zigzag-oriented edge (dotted white lines) with larger scale (~10 nm) roughness making many 120° turns. Inset (scale bar =4.3 Å) shows representative atomically resolved image of graphene lattice inside the grain (I=20 nA, V=-0.2 V). The asymmetry between A and B sublattices seen in the image and some other areas of graphene (eg. Fig. 1 c,d) might be related with the tunneling condition used and graphene interaction with the underlying substrate lattice.

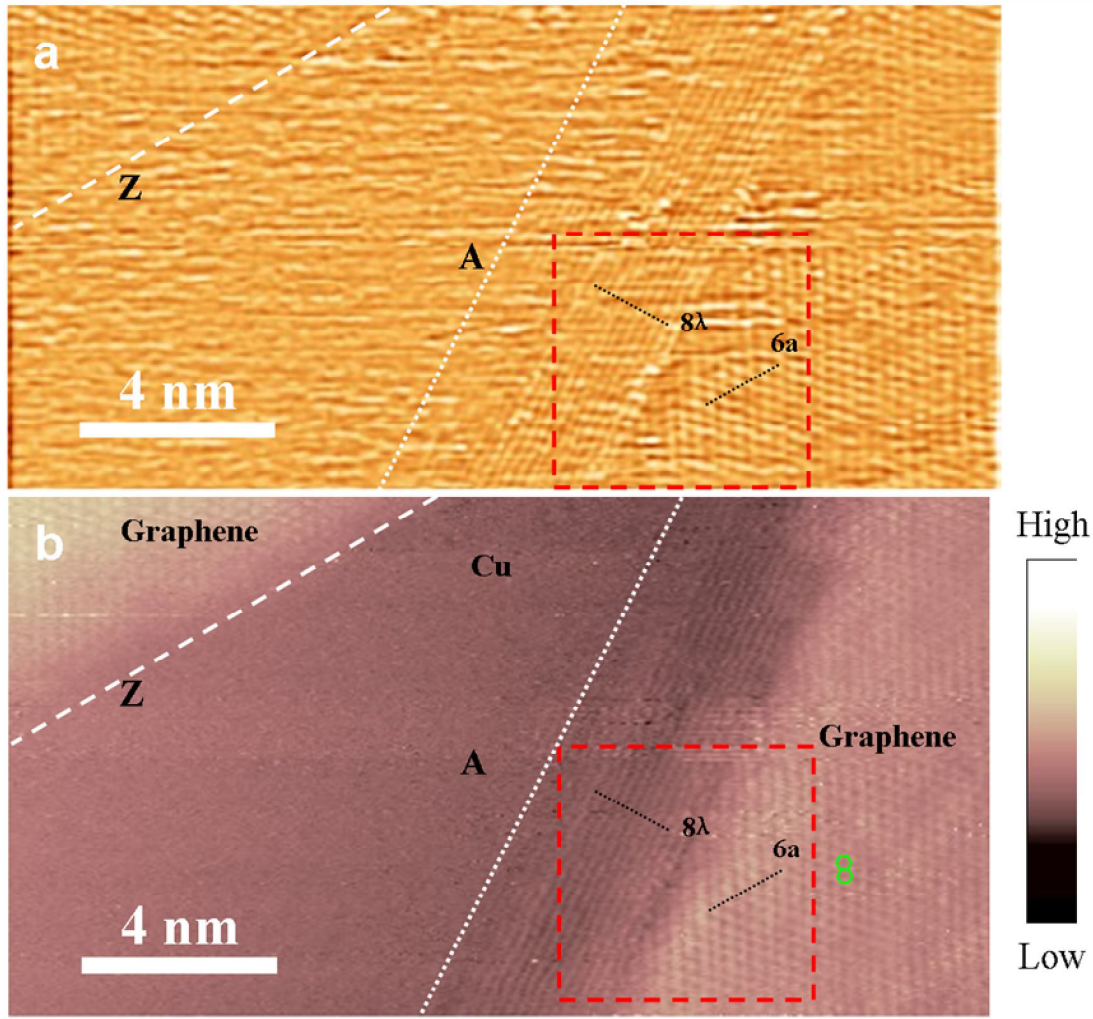

**Figure S2** Atomically-resolved topographic image (a) (processed with a wavelet-based filter[S2] to enhance the contrast.) and its corresponding differential conductance dI/dV image (b) recorded simultaneously (I=10 nA, V=0.2 V, acquired from the dotted green box in Fig. 2A), containing the rare-occurring armchair edge and another nearby zigzag edge. Dotted and dashed white lines mark the armchair ("A") and zigzag ("Z") directions near the corresponding edges. The standing wave pattern parallel to the armchair edge is readily seen. The two black dotted lines of equal length measure approximately 8 oscillation periods ($\lambda$) of the wave pattern and 6 units of graphene lattice constants ($a$). Fig. 2c is resized from the boxed area (red dashed lines) to partially correct a small distortion in this image (Fig. S2a) resulted from a hysteresis in STM tip movement and sample/tip drift. The correction for Fig. 2c used the expected graphene honeycomb lattice inside the grain as a guide and has no appreciable effect on the $\lambda/a$ value extracted.

17